\documentclass[aps,prfluids,preprint,superscriptaddress,amsmath]{revtex4-2}

\usepackage{amsmath,amssymb,amsthm}
\usepackage{graphicx}
\usepackage{multirow}
\usepackage{subcaption}
\usepackage{hyperref}

\begin{document}

\title{Stokes flows in a two-dimensional bifurcation}

\author{Yidan Xue}
\email{xue@maths.ox.ac.uk}
\affiliation{Mathematical Institute, University of Oxford, UK}
\affiliation{School of Mathematics, Cardiff University, UK}

\author{Stephen J. Payne}
\email{stephenpayne@ntu.edu.tw}
\affiliation{Institute of Applied Mechanics, National Taiwan University, Taiwan}

\author{Sarah L. Waters}
\email{waters@maths.ox.ac.uk}
\affiliation{Mathematical Institute, University of Oxford, UK}



\date{\today}

\begin{abstract}
The flow network model is an established approach to approximate pressure-flow relationships in a bifurcating network, and has been widely used in many contexts. Existing models typically assume unidirectional flow and exploit Poiseuille's law, and thus neglect the impact of bifurcation geometry and finite-sized objects on the flow. We determine the impact of bifurcation geometry and objects by computing Stokes flows in a two-dimensional (2D) bifurcation using the LARS (Lightning-AAA Rational Stokes) algorithm, a novel mesh-free algorithm for solving 2D Stokes flow problems utilising an applied complex analysis approach based on rational approximation of the Goursat functions. We compute the flow conductances of bifurcations with different channel widths, bifurcation angles, curved boundary geometries, and fixed circular objects. We quantify the difference between the computed conductances and their Poiseuille's law approximations to demonstrate the importance of incorporating detailed bifurcation geometry into existing flow network models. We parameterise the flow conductances of 2D bifurcation as functions of the dimensionless parameters of bifurcation geometry and a fixed object using a machine learning approach, which is simple to use and provides more accurate approximations than Poiseuille's law. Finally, the details of the 2D Stokes flows in bifurcations are presented.
\end{abstract}

\maketitle
\newpage

\section{Introduction}

The computation of fluid flows within networks underpins many biological, industrial and engineering applications. For instance, quantifying blood flow in an organ or an organ system can help us better understand their physiological functions \cite{Fung2013}. Other biological applications include microvascular flows \cite{Secomb2017,El-Bouri2021}, oxygen transport \cite{Secomb2004,Hartung2021,Xue2021,Xue2022}, drug delivery \cite{desposito2018}, and microfluidic devices \cite{Oh2012,Merlo2022}. Flow networks have also been used to estimate permeability and multiphase flow properties in porous media \cite{Blunt2002,Baychev2019}.

The standard way to approximate flows in a network is via a flow network modelling approach, which is a zero-dimensional (0D) model of the underlying three-dimensional (3D) flow problem, where junctions and boundary points of the network are represented by nodes with flow segments between them. By assuming steady unidirectional flow subject to no-slip wall conditions in each flow segment, the flow conductance is then described by Poiseuille's law \cite{Sutera1993}. For a flow segment modelled as a tube of circular cross-section, the conductance is
\begin{equation}
G=\frac{q}{\Delta{p}}=\frac{\pi{d}^4}{128\mu{l}},\label{eq:poiseuille}
\end{equation}
where $q$ is the flow rate, $\Delta{p}$ is the pressure drop across the segment, $d$ is the diameter, $\mu$ is the viscosity (a constant or a function of diameter), and $l$ is the segment length. Poiseuille's law applies to Newtonian viscous fluid or a non-Newtonian fluid that can be represented as a Newtonian fluid with an effective viscosity. Note that Poiseuille's law for two-dimensional (2D) channels is $G=q/\Delta{p}=d^3/12{\mu}l$, where now $d$ is the channel width. 

By imposing continuity of flux at internal nodes, a linear system for pressures and fluxes can be constructed via
\begin{equation}
\mathbf{G}\mathbf{P}=\mathbf{Q},\label{eq:linear}
\end{equation}
where $\mathbf{G}$ is the conductance tensor, $\mathbf{P}$ is the vector of nodal pressures, and $\mathbf{Q}$ is the vector of segment fluxes. The conductance tensor is a property of the network geometry and provides a simplified representation of the relationship between pressures and fluxes. Note that the conductance tensor is also affected by the presence of finite-sized objects in the network. In addition, the pressure-flux relationship is no longer linear when the viscosity depends on factors such as the volume of red blood cells when considering blood flow \cite{Pries1992,Pop2007,Secomb2017}. The flow and red blood cell distribution are then computed via an iterative solver \cite{Gould2015}.

While Poiseuille's law is based on the assumption of unidirectional flow, this assumption no longer holds at junction regions or near finite-sized objects, and the Poiseuille's law approximation for flow conductance in networks loses accuracy. Furthermore, detailed flow modelling becomes extremely useful for understanding local stress distributions for tissue growth \cite{O'Dea2010}, advective transport of finite-sized objects \cite{Audet1987}, or advection-diffusion of tiny particles \cite{Yeo2021}.

In this paper, we use a 2D Stokes flow model to propose an updated flow conductance model for a 2D bifurcation, which considers both the bifurcation geometry and the presence of finite-sized objects, and thus accurately predicts the pressure-flux relationship. We note that the same model reduction concept can be used to consider conductances in a 3D bifurcation. Additionally, we use the Stokes flow model to examine the details of 2D flow in bifurcations.

To the best of our knowledge, Stokes flows in bifurcations have not yet been thoroughly studied either analytically or numerically, even for a 2D setup. Analytical or semi-analytical solutions exist for simpler 2D Stokes flow problems, including flows near a corner \cite{Moffatt1964}, in a partitioned channel \cite{Jeong2001}, in an expanded channel \cite{Luca2018} and in a constricted channel \cite{Tavakol2017}. However, these analytical techniques are not suitable for solving 2D Stokes flows in an bifurcation with complex boundary geometries. Alternatively, numerical methods can be used to compute 2D Stokes flows in a bifurcation, including finite element methods \cite{Logg2012} and boundary integral methods \cite{Pozrikidis1992}. However, these methods are computationally expensive, which may prohibit a comprehensive interrogation of parameter space, even for 2D scenarios.

Recent developments in rational approximation \cite{Nakatsukasa2018,Gopal2019,Brubeck2021,Costa2023} have underpinned novel algorithms to compute 2D Stokes flows in general domains using an applied complex variable approach \cite{Brubeck2022,Xue2024}. Following the lightning algorithm for solving Laplace's equation \cite{Gopal2019}, Brubeck and Trefethen \cite{Brubeck2022} developed the lightning Stokes solver. The stream function, which satisfies the biharmonic equation, was represented using two complex analytic Goursat functions \cite{Goursat1898} and approximated by rational functions with clustering poles distributed exponentially near corners \cite{Brubeck2022}. Based on the lightning algorithm, Xue et al.~developed the LARS (Lightning-AAA Rational Stokes) algorithm \cite{Xue2024} for computing 2D Stokes flows in general domains, including domains which have curved boundaries or are multiply connected. The computations typically take less than one second on a laptop and gives solutions with at least 6-digit accuracy \cite{Xue2024}. 

In this paper, we use the LARS algorithm to compute Stokes flows in 2D bifurcations considering different bifurcation and particle geometries. Our computations encompass bifurcations with different channel widths, bifurcation angles, curved boundary geometries, as well as the scenarios with a fixed circular particle of varying size and location. We compute the flow conductances in three channels, and compare these against their Poiseuille's law approximations to demonstrate the need to incorporate detailed flow modelling into flow network models. We then parameterise flow conductances as functions of dimensionless geometrical parameters of the bifurcation using a machine learning approach, facilitating their applications to 2D flow network simulations. Furthermore, we present flow features that cannot be captured by a network model, including streamline patterns and flow separations \cite{Dean1949,Moffatt1964,Michael1977}.

In Section \ref{sec:problem}, we formulate the physical problem in dimensionless form, and give a representation for the flow conductances of a 2D bifurcation. In Section \ref{sec:lars}, we introduce the LARS algorithm for computing 2D Stokes flows using rational approximation. The results are presented in Section \ref{sec:results}, followed by discussions in Section \ref{sec:discussion}.

\section{Problem formulation}
\label{sec:problem}

\begin{figure}
  \centering
  \begin{subfigure}{.45\textwidth}
  \centering
  \includegraphics[width=\textwidth]{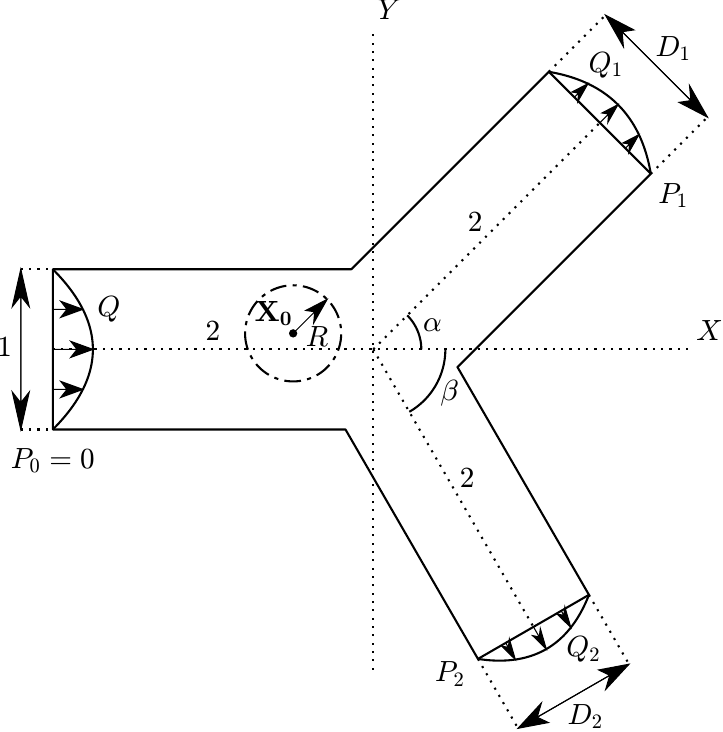}
  \caption{2D Stokes flow}
\end{subfigure}%
\begin{subfigure}{.45\textwidth}
  \centering
  \includegraphics[width=\textwidth]{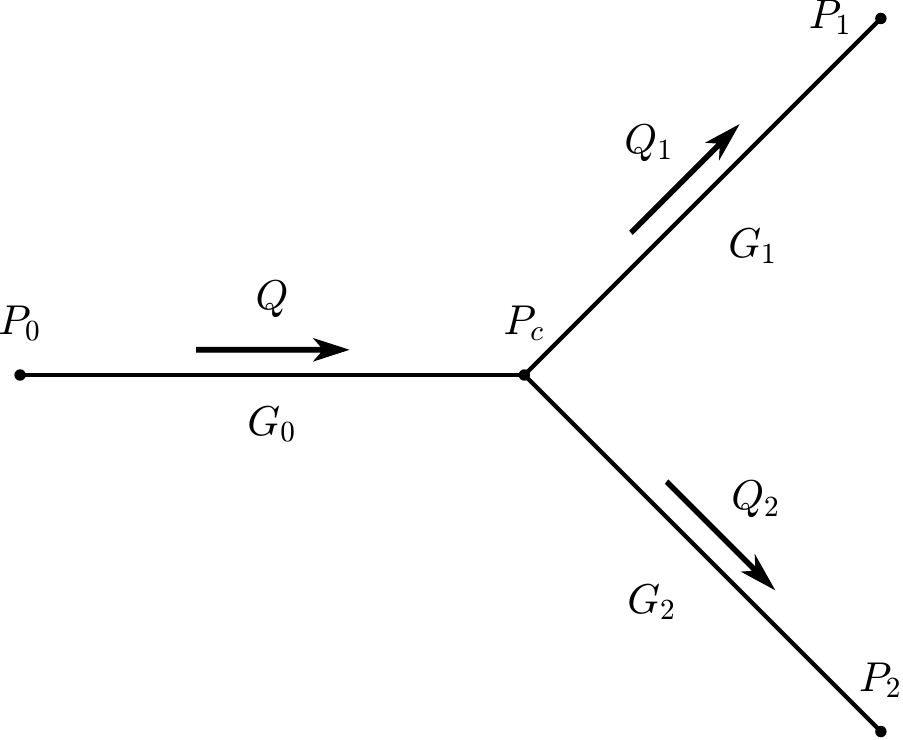}
  \vspace{6pt}
  \caption{0D Flow network}
\end{subfigure}
  \caption{Schematic of Stokes flow through a 2D bifurcation (a) and its flow network representation (b). The solid and dash-dotted lines in (a) indicate the domain and particle boundaries (Section \ref{sec:object}), respectively.}
  \label{fig:bifurcation}
\end{figure}

We consider a 2D bifurcation with one inlet parent channel with width $d$, and two outlet child channels with widths $d_1$ and $d_2$, respectively, as shown in Fig.~\ref{fig:bifurcation}(a). The centrelines of three channels intersect at the origin of a Cartesian coordinate system $\textbf{x}=(x,y)^T$. The angles between the positive $x$ axis and the two channel centrelines are denoted $\alpha$ and $\beta$, respectively. Each channel has centreline length $l$.

We consider steady flow of an incompressible Newtonian viscous fluid governed by the Stokes and continuity equations:
\begin{equation}
    \mu\nabla^2\mathbf{u}=\nabla{p},\ 
    \nabla\cdot\mathbf{u}=0, \label{eq:stokes}
\end{equation}
where $\mathbf{u}=(u,v)^T$ is the 2D velocity field, $p$ is the pressure, and $\mu$ is the viscosity. We prescribe normal stress and parallel flow boundary conditions at the three channel openings, and zero velocity boundary conditions on the channel walls. The boundary conditions are equivalent to flow entering and exiting the bifurcation with a fully-developed parabolic flow profile. The flow is driven by the pressure values prescribed at the channel openings. The flux in the inlet parent channel is then $q=\int_\Gamma\mathbf{u}\cdot\mathbf{n}ds$, where $\Gamma$ is a channel cross-section and $\mathbf{n}$ is the unit normal vector. The fluxes in two outlet child channels are $q_1$ and $q_2$, respectively.

\subsection{Non-dimensionalisation}

We non-dimensionalise as follows
\begin{equation}
\textbf{X} = \frac{\textbf{x}}{d},\ \textbf{U}=\frac{\textbf{u}}{q/d}, P = \frac{p}{\mu{q}/d^2},
\end{equation}
where capitals denote dimensionless variables. The dimensionless Stokes equations in component form, together with the continuity equation, then become
\begin{align}
\frac{\partial^2U}{\partial{X}^2}+\frac{\partial^2U}{\partial{Y}^2}&=\frac{\partial{P}}{\partial{X}}, \label{eq:nd1}\\
\frac{\partial^2V}{\partial{X}^2}+\frac{\partial^2V}{\partial{Y}^2}&=\frac{\partial{P}}{\partial{Y}}, \label{eq:nd2}\\
\frac{\partial{U}}{\partial{X}}+\frac{\partial{V}}{\partial{Y}}&=0. \label{eq:nd3}
\end{align}
We set the dimensionless centreline length $L=l/d=2$ for each channel to ensure the flow domain has sufficient length to achieve fully-developed flow at the outlets. Without loss of generality, we set the dimensionless pressure at the inlet as $P_0=0$. For a 2D bifurcation, the system is then characterised by the following dimensionless parameters: child channel widths $D_1=d_1/d$ and $D_2=d_2/d$, outlet pressures $P_1=p_1d^2/\mu{q}$ and $P_2=p_2d^2/\mu{q}$, and bifurcation angles $\alpha$ and $\beta$. 

In Section \ref{sec:boundary_geometry}, we compute the flow in bifurcations with curved boundary geometries, where the channel boundaries are described by cubic B\'{e}zier curves. In Section \ref{sec:object}, we consider flow past a fixed cylindrical particle in the bifurcation. The dimensionless centre location and radius of the particle are $\mathbf{{X}_0}=(X_0,Y_0)$ and $R$, respectively. 

Having defined the dimensionless problem, we now formulate a linear relationship between the pressures ($P_1$ and $P_2$) and fluxes ($Q_1=q_1/q$ and $Q_2=q_2/q$) at the two outlets.

\subsection{0D network model}

The relationship between pressures and fluxes of Stokes flows in the bifurcation can be reduced to a 0D network model (Fig.~\ref{fig:bifurcation}(b)). The dimensionless fluxes in the flow segments are related via
\begin{equation}
Q = Q_1+Q_2 = G_0(P_0-P_c),\ 
Q_1 = G_1(P_c-P_1),\ 
Q_2 = G_2(P_c-P_2), \label{eq:network}
\end{equation}
where $G_0$, $G_1$ and $G_2$ are flow conductances that depend on the bifurcation and particle geometry. When $P_0=0$,
\begin{equation}
P_c = \frac{G_1P_1+G_2P_2}{G_0+G_1+G_2}.
\end{equation}
From (\ref{eq:network}), we now have a linear system relating the pressures and fluxes at two outlets, which we can represent by (\ref{eq:linear}), where $\mathbf{P}=(P_1,P_2)^T$, $\mathbf{Q}=(Q_1,Q_2)^T$ and 
\begin{equation}
\mathbf{G}=\begin{bmatrix}
\dfrac{-G_1G_2-G_1G_0}{G_0+G_1+G_2} & \dfrac{G_1G_2}{G_0+G_1+G_2}\\[1em]
\dfrac{G_1G_2}{G_0+G_1+G_2} & \dfrac{-G_2G_1-G_2G_0}{G_0+G_1+G_2}
\label{eq:tensor}
\end{bmatrix}
\end{equation}
is the conductance tensor.

When two solutions of flux vector $\mathbf{Q}$ for two linearly independent $\mathbf{P}$ (e.g.~$\mathbf{P}=[1\ 0]^T$ and $\mathbf{P}=[0\ 1]^T$) are provided, the rank 2 conductance tensor $\mathbf{G}$ and the three components $G_0$, $G_1$ and $G_2$ can be calculated. Using $\mathbf{G}$ and (\ref{eq:linear}), we can then predict flux for any set of pressure conditions.

For an idealised bifurcation, where the junction is simplified as a node and the fluid flow in each channel is assumed to be fully-developed, we can approximate the three flow conductances using Poiseuille's law for 2D channel flows \cite{Sutera1993}:
\begin{equation}
\tilde{G}_0=\dfrac{1}{12L},\ 
\tilde{G}_1=\dfrac{D_1^3}{12L},\ 
\tilde{G}_2=\dfrac{D_2^3}{12L},\label{eq:poiseuille_approx}
\end{equation}
where $\tilde{G}$ is the idealised flow conductance. Using (\ref{eq:tensor}) and (\ref{eq:poiseuille_approx}), the pressure-flux relationship for the idealised 2D bifurcation can be approximated by
\begin{equation}
\begin{bmatrix}
\dfrac{-D_1^3D_2^3-D_1^3}{12L(1+D_1^3+D_2^3)} & \dfrac{D_1^3D_2^3}{12L(1+D_1^3+D_2^3)}\\[1em]
\dfrac{D_1^3D_2^3}{12L(1+D_1^3+D_2^3)} & \dfrac{-D_1^3D_2^3-D_2^3}{12L(1+D_1^3+D_2^3)}
\end{bmatrix}
\mathbf{P}=\mathbf{\tilde{G}}\mathbf{P}=\mathbf{Q}.\label{eq:Poiseuille_tensor}
\end{equation}

In this paper, we will compute $\mathbf{G}$ for bifurcations with different geometries using the numerical method presented in the next section, and compare those against $\mathbf{\tilde{G}}$ to show the difference.

\section{Computing 2D Stokes flows via rational approximation}
\label{sec:lars}

We compute the 2D Stokes flows in a bifurcation using the LARS algorithm \cite{Xue2024}, which uses an applied complex variable approach with rational approximation. The LARS algorithm results in at least 6-digit accuracy in less than 1 second. In this section, we summarise the LARS algorithm, referring the interested reader to further details in \cite{Xue2024}.

\subsection{The Goursat representation for biharmonic equations}

We define a stream function $\psi$ for the dimensionless Stokes flow problem as
\begin{equation}
    U=\frac{\partial\psi}{\partial{Y}},\ V=-\frac{\partial\psi}{\partial{X}},
\end{equation}
which satisfies the biharmonic equation
\begin{equation}
\nabla^4\psi=0.
\end{equation}
In the complex plane $\zeta=X+iY$, where $i=\sqrt{-1}$, we have
\begin{equation}
    \frac{\partial}{\partial{\zeta}}=\frac{1}{2}\left(\frac{\partial}{\partial{X}}-i\frac{\partial}{\partial{Y}}\right),\ \frac{\partial}{\partial{\bar{\zeta}}}=\frac{1}{2}\left(\frac{\partial}{\partial{X}}+i\frac{\partial}{\partial{Y}}\right),
\end{equation}
where overbars denote complex conjugates. Then the biharmonic equation can be rewritten as
\begin{equation}
\frac{\partial^4\psi}{\partial^2\zeta\partial^2\bar{\zeta}}=0,
\end{equation}
which has a solution in the form
\begin{equation}
\psi(\zeta,\bar{\zeta})=\mathrm{Im}[\bar{\zeta}f(\zeta)+g(\zeta)],
\end{equation}
where $f(\zeta)$ and $g(\zeta)$ are two analytic functions in the fluid domain, known as the Goursat functions \cite{Goursat1898}. The dimensionless velocity, pressure and vorticity magnitude ($\Omega = \partial{V}/\partial{X}-\partial{U}/\partial{Y}$) can then be expressed in terms of Goursat functions as
\begin{align}
U-iV = -\overline{f(\zeta)}+\bar{\zeta}f'(\zeta)+g'(\zeta), \label{eq:bc1} \\
P-i\Omega=4f'(\zeta). \label{eq:bc2}
\end{align}
The Goursat representation satisfies the biharmonic equation by construction. To solve the Stokes flow problem, we therefore need to determine the Goursat functions such that the boundary conditions are satisfied.

\subsection{Approximating the Goursat functions using rational functions}

The LARS algorithm approximates the Goursat functions using rational functions \cite{Brubeck2022,Xue2024}. The rational function basis consists of poles for the singularities \cite{Nakatsukasa2018,Gopal2019,Brubeck2022}, polynomial for the smooth part \cite{Brubeck2021}, and Laurent series for multiply connected domains \cite{Price2003,Trefethen2018} of the solution. This leads to a rational function $r(\zeta)$ in the form:
\begin{equation}
r(\zeta)=\sum_{j=1}^m\frac{a_j}{\zeta-z_j}+\sum_{j=0}^n{b_j}\zeta^j+\sum_{i=1}^p\sum_{j=1}^q{c_{ij}}(\zeta-\zeta_i)^{-j},
\label{eq:rational}
\end{equation}
where $a_j$, $b_j$ and $c_{ij}$ are complex coefficients to be determined, $z$ are poles and $\zeta_i$ is a point in the $i$th hole (recalling the particle in Fig.~\ref{fig:bifurcation}). Note that we also need to include two log terms in two Goursat functions (one for $f(\zeta)$ and one for $g(\zeta)$) corresponding to each hole, due to the logarithmic conjugation theorem \cite{Axler1986,Price2003}.

It has been shown that an analytic function in a polygon domain can be approximated with a root-exponential convergence, if the poles are exponentially clustered near corner singularities \cite{Gopal2019}. This leads to the lightning algorithm for computing Laplace problems \cite{Gopal2019} and 2D Stokes flows \cite{Brubeck2022} in polygon domains. For bifurcations with sharp corners, we follow the lightning algorithm to place poles clustering towards each corner singularity.

For bifurcations with smooth boundaries, we first approximate the Schwarz function $F(\zeta)=\bar{\zeta}$ on each curved boundary using the AAA algorithm \cite{Nakatsukasa2018,Costa2023}. The AAA algorithm searches for the best rational approximation in a barycentric form automatically. We choose the Schwarz function here, because it only depends on the boundary shape instead of the boundary value. After finding the rational function that approximates the boundary shape, we use its poles outside the domain to approximate the Goursat functions. This is known as the AAA-least squares approximation \cite{Costa2023}.

We perform a Vandermonde with Arnoldi orthogonalization \cite{Brubeck2021} to construct a well-conditioned basis for the polynomial, poles and Laurent series. Imposing the boundary conditions of the fluid problem (\ref{eq:bc1}, \ref{eq:bc2}), the complex coefficients $a$, $b$ and $c$ in two Goursat functions can be computed easily by solving a least-squares problem. The computation is carried out using MATLAB and all codes are available in the GitHub repository provided in Appendix \ref{sec:appendix}.

\section{Results}
\label{sec:results}

Using the LARS algorithm, we present Stokes flows in 2D bifurcations with different geometries including child channel widths, bifurcation angles and curved boundary geometries. We also consider bifurcations with circular particles inside and quantify their impact on the pressure-flux relationship. In Sections \ref{sec:bifurcation_example}-\ref{sec:object}, we compute the flow conductances for different bifurcation geometries, and compare these against the Poiseuille's law approximation (\ref{eq:Poiseuille_tensor}) where applicable. In Section \ref{sec:streamline}, we investigate details of the 2D flow in bifurcations, which cannot be captured using a flow network model.

\subsection{2D Stokes flows in a bifurcation}
\label{sec:bifurcation_example}

\begin{figure}
  \centering
  \includegraphics[width=.6\textwidth]{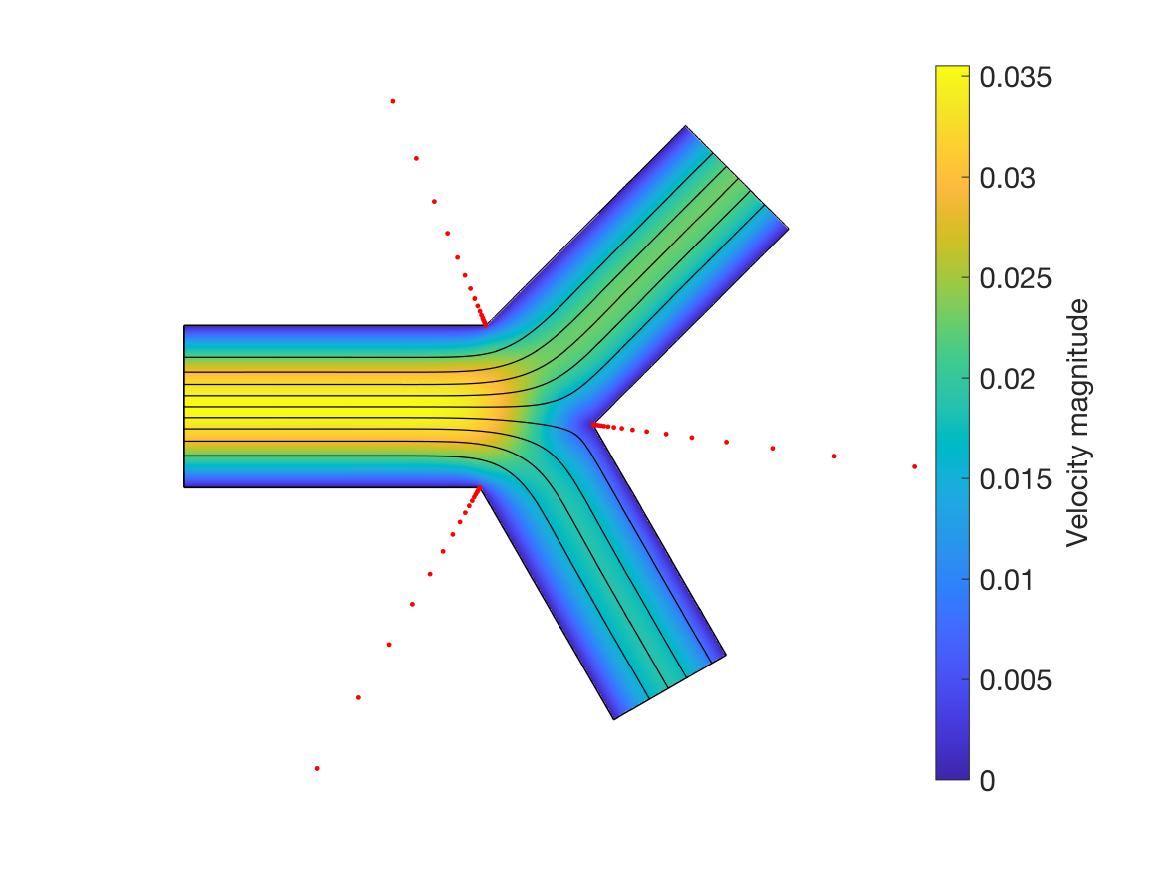}
  \caption{Stokes flows in a 2D bifurcation for $P_1=P_2=-1$, solved by the lightning algorithm, where $D_1=0.9$, $D_2=0.8$, $\alpha=\pi/4$, $\beta=\pi/3$ and $L=2$. The streamlines are denoted by black lines and the velocity magnitude is represented by a colourmap. The locations of the poles are marked by red dots.}
  \label{fig:bifurcation_example}
\end{figure}

Figure \ref{fig:bifurcation_example} presents the Stokes flows in a typical 2D bifurcation, where $D_1=0.9$, $D_2=0.8$, $\alpha=\pi/4$, $\beta=\pi/3$ and $L=2$ for $P_1=P_2=-1$. The lightning algorithm places poles exponentially clustered towards three sharp corners of the geometry, where the placement of poles are indicated by red dots in Fig.~\ref{fig:bifurcation_example}. Using a polynomial of degree 24 with 48 poles clustered near each corner, a solution can be computed in less than 0.5 second. The maximum error in dimensionless velocity components and pressures on the domain boundary is less than $10^{-8}$. For this geometry, the computed conductances are $G_0=0.0422$, $G_1=0.0313$, and $G_2=0.0226$, while the idealised conductances are $\tilde{G}_0=0.0417$, $\tilde{G}_1=0.0304$, and $\tilde{G}_2=0.0213$ with relative percentage errors 1.34\%, 2.93\%, and 5.73\%, respectively.

\subsection{Effects of channel width on conductance}
\label{sec:diameter_effects}

\begin{figure}
  \centering
  \includegraphics[width=\textwidth]{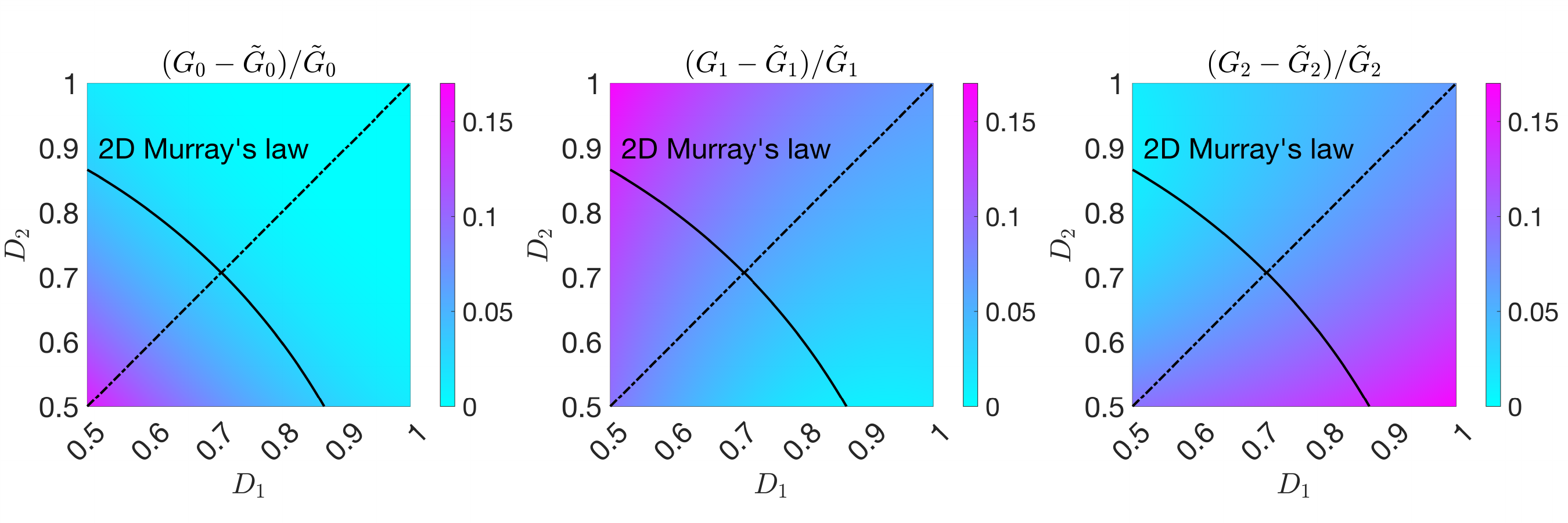}
  \caption{Relative differences in $G_0$, $G_1$ and $G_2$ from 2D Stokes flows simulations, compared with Poiseuille's law approximations for $D_1,D_2\in[0.5,1]$, when $\alpha=\pi/4$, $\beta=\pi/4$ and $L=2$. The black curve indicates the possible widths of two child channels based on Murray's law in 2D. The symmetry of the parameter space in each figure is indicated by a dash-dotted line.}
  \label{fig:diameter_effects}
\end{figure}

For a bifurcation consisting of 3 straight channels with $L=2$, the geometry has four degrees of freedom: $D_1$, $D_2$, $\alpha$ and $\beta$. To investigate the effects of the width of two child channels on the computed conductances, we first set the bifurcation angles $\alpha=\pi/4$ and $\beta=\pi/4$ (we investigate the effects of changing the bifurcation angles in Section \ref{sec:angle}). Figure \ref{fig:diameter_effects} shows the relative differences between the computed conductances $G_0$, $G_1$ and $G_2$ and their Poiseuille's law approximations, for $D_1,D_2\in[0.5,1]$. The black curve indicates the possible widths of the two child channels if they obey Murray's law \cite{Murray1926} in 2D: $D_1^2+D_2^2=D^2=1$.
Note that every term is to the third power in the original Murray's law for 3D problems \cite{Murray1926}.

For the considered parameters, we see the Poiseuille's law approximation underestimates the flow conductance in two child channels. The error in the flow conductance in the parent channel only becomes significant when $D_1$ and $D_2$ are close to 0.5. Furthermore, $(G_1-\tilde{G}_1)/\tilde{G}_1$ is 16.6\%, while $(G_2-\tilde{G}_2)/\tilde{G}_2$ is approximately 1\%, when $D_1=0.5$ and $D_2=1$. This indicates that the Poiseuille's law approximation underestimates not only the total flux through a bifurcation, but also the fraction of flux that enters the first channel, when $D_1=0.5$ and $D_2=1$. Note that these values are for $L=2$. For bifurcations with larger channel lengths, the mismatch between the two conductances will be reduced, since Poiseuille's law becomes a more accurate approximation for the conductance of fully-developed flows in straight channels.

\subsection{Effects of bifurcation angle on flow partition}
\label{sec:angle}

\begin{figure}
  \centering
  \includegraphics[width=.8\textwidth]{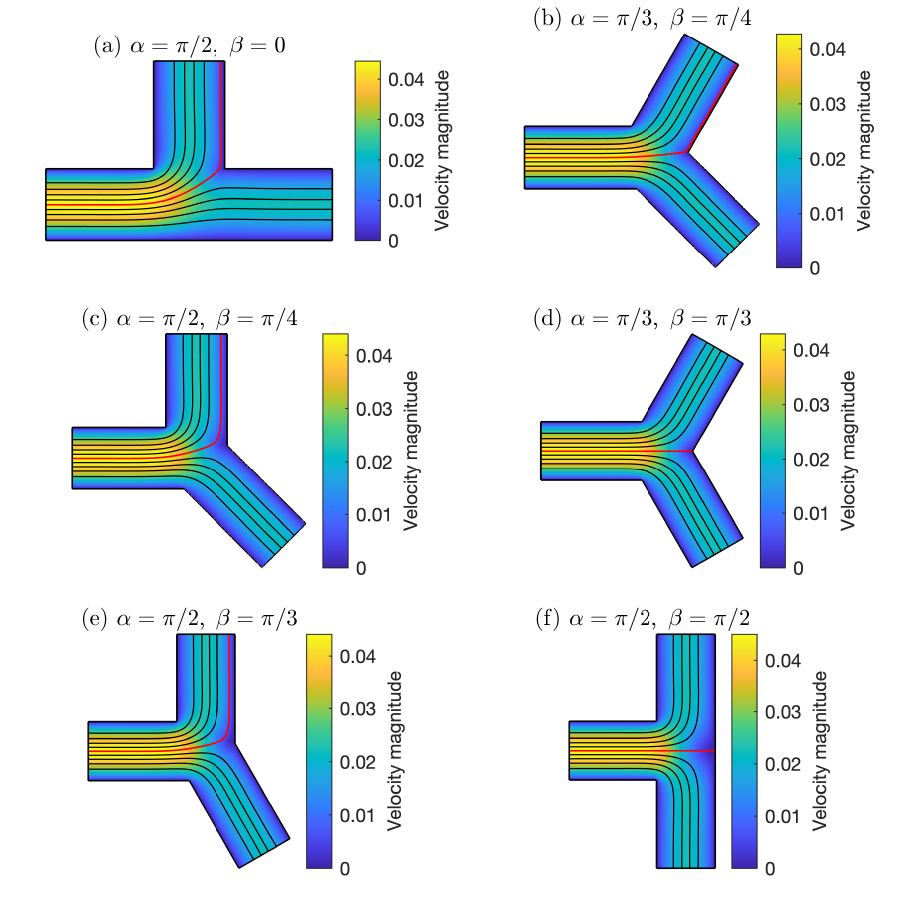}
  \caption{Stokes flows in a 2D bifurcation with different bifurcation angles, when $D_1=D_2=1$, $L=2$ and $P_1=P_2=-1$. The centre streamline of the parent channel is coloured in red. Other streamlines are coloured in black.}
  \label{fig:angle_cases}
\end{figure}

Figure \ref{fig:angle_cases} presents Stokes flows in bifurcations with different bifurcation angles, when $D_1=D_2=1$, $L=2$ and $P_1=P_2=-1$. Poiseuille's law predicts that the child channels will have an even flow partition regardless of bifurcation angle. However, Stokes flow simulations reveal that the first channel receives more flow than the second channel in cases (a), (b), (c) and (e). We observe that the centre streamline of the parent channel (coloured in red), which is the streamline corresponding to equal flow partition, enters the first channel in these cases. 

\begin{figure}
  \centering
  \includegraphics[width=\textwidth]{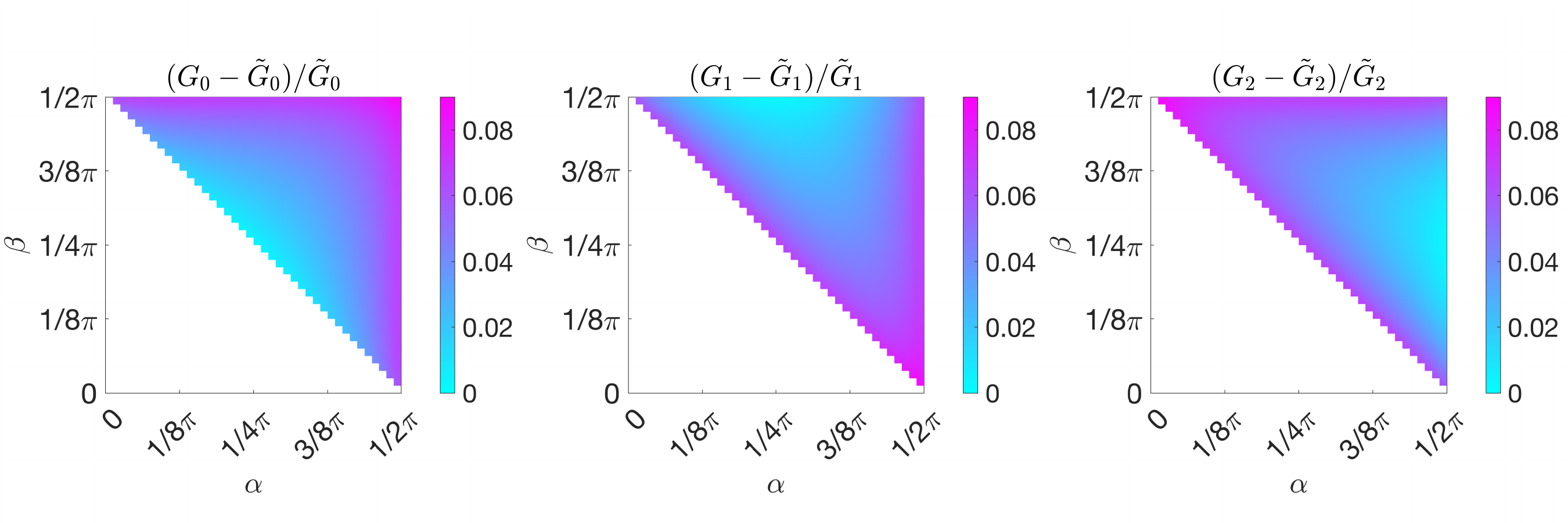}
  \caption{Relative differences in $G_0$, $G_1$ and $G_2$ from 2D Stokes flows simulations, compared with Poiseuille's law approximations for different bifurcation angles, when $D_1=D_2=1$ and $L=2$.}
  \label{fig:angle_effects}
\end{figure}

While setting $D_1=D_2=1$ and $L=2$, we perform a parameter sweep for $\alpha,\beta\in[0,\pi/2]$ except when $|\alpha+\beta|<\pi/2$, since this will lead to a small angle or an overlap between the two child channels. As shown in Fig.~\ref{fig:angle_effects}, Poiseuille's law underestimates the flow conductance in the three flow segments, while the maximum errors in $G_0$, $G_1$ and $G_2$ happen at $\alpha=\beta=\pi/2$; $\alpha=\pi/2$ and $\beta=0$; and $\alpha=0$ and $\beta=\pi/2$, respectively. When $\alpha=\pi/2$ and $\beta=0$, the computed conductance in the first channel is about 9\% higher than the idealised conductance, while the difference in the other channel is much smaller. This leads to more flow in the first channel than the second channel as shown in cases (a), (c) and (e) in Fig.~\ref{fig:angle_cases}, where $\alpha=\pi/2$ and $\beta<\pi/2$.

\subsection{Effects of curved boundary geometry on bifurcation flow}
\label{sec:boundary_geometry}

\begin{figure}
  \centering
  	\begin{subfigure}{0.49\textwidth}
		\centering
		\includegraphics[width=\textwidth]{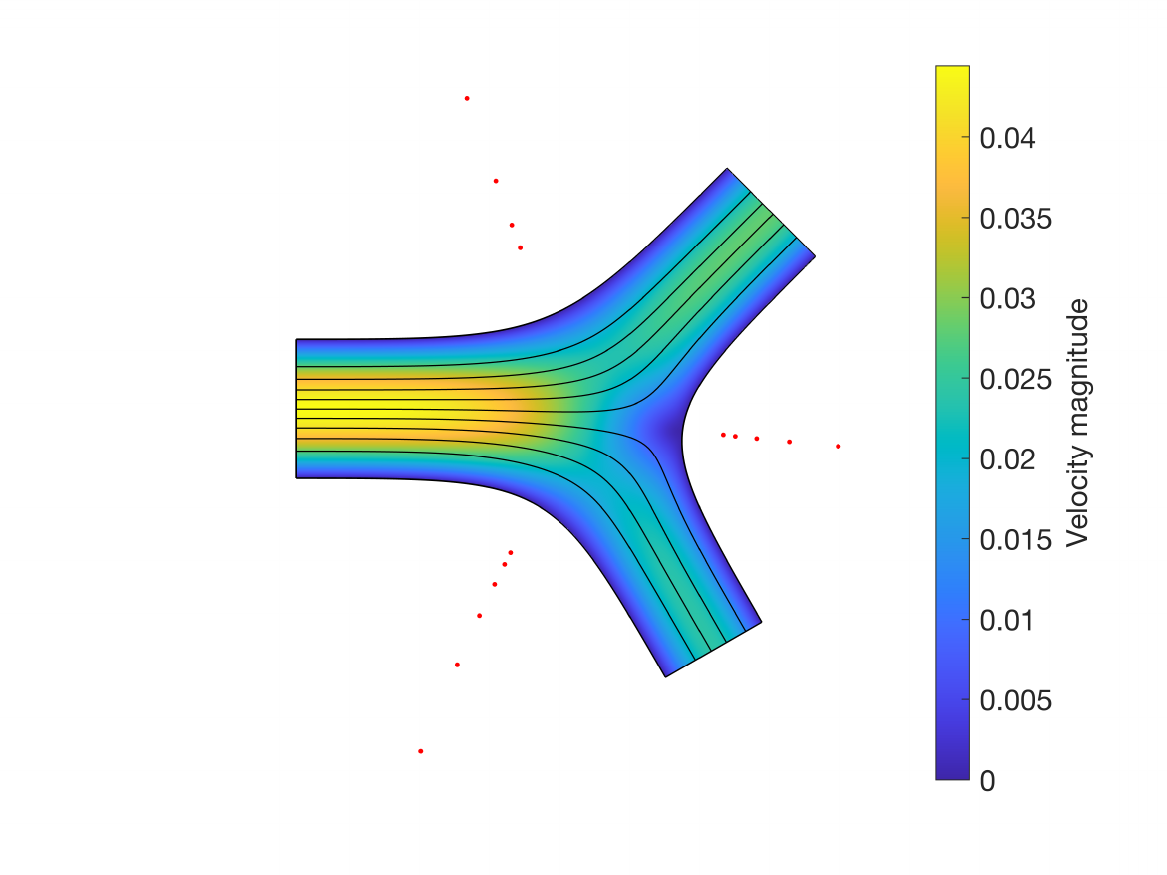}
		\caption{Simulation results}
	\end{subfigure}
	\begin{subfigure}{0.49\textwidth}
	    \centering
		\includegraphics[width=.55\textwidth]
  {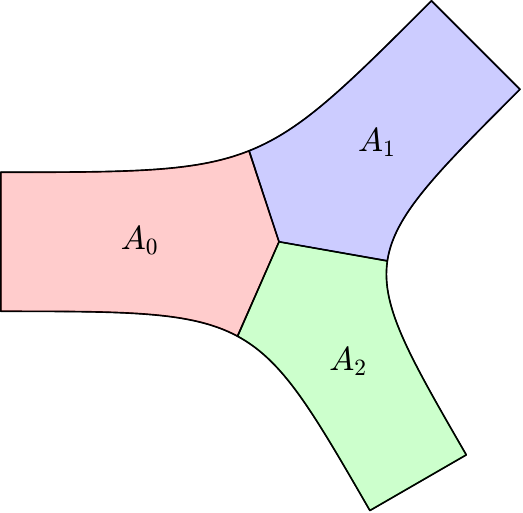}
  \vspace{18pt}
		\caption{Area for each channel}
	\end{subfigure}
  \caption{(a) Stokes flows in a 2D bifurcation with smooth boundaries solved by the LARS algorithm, where $D_1=0.9$, $D_2=0.8$, $\alpha=\pi/4$, $\beta=\pi/3$, $L=2$ and $P_1=P_2=-1$. The streamlines are denoted by black lines and the velocity magnitude is represented by a colourmap. The AAA poles are marked by red dots. (b) Area for each channel when calculating area preserved conductance $\hat{G}$.}
  \label{fig:smooth_example}
\end{figure}

Figure \ref{fig:smooth_example}(a) presents the Stokes flows computed in the same bifurcation as shown in Fig.~\ref{fig:bifurcation}, but with curved boundaries. Using a polynomial of degree 48 with poles placed by the AAA algorithm \cite{Nakatsukasa2018,Costa2023} near the three curved boundaries, the solution is computed to the accuracy $\mathcal{O}(10^{-7})$.

\begin{figure}
  \centering
  \includegraphics[width=\textwidth]{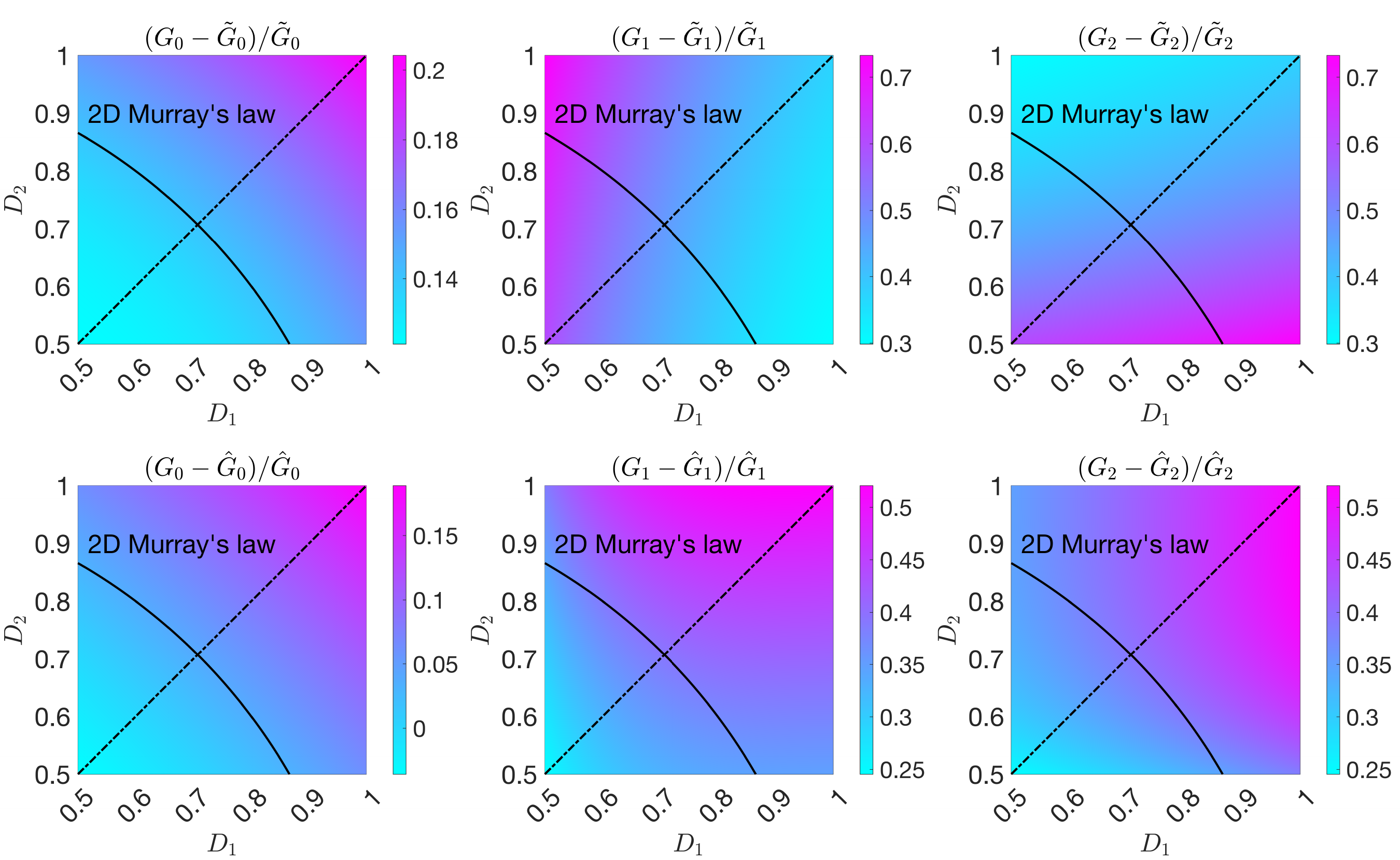}
  \caption{Relative differences in $G_0$, $G_1$ and $G_2$, compared with their Poiseuille's law approximations based on inlet and outlet channel width (top row), or scaled channel width to preserve bifurcation area (bottom row), for the bifurcation with smooth boundaries ($D_1,D_2\in[0.5,1]$, when $\alpha=\pi/4$, $\beta=\pi/4$ and $L=2$). The black curve indicates the possible widths of two child channels based on Murray's law in 2D. The symmetry of the parameter space in each figure is indicated by a dash-dotted line. Note that the colour scale varies for each figure.}
  \label{fig:diameter_effects_smooth}
\end{figure}

In the same parameter space as Section \ref{sec:diameter_effects}, we compute $G_0$, $G_1$ and $G_2$ using 2D Stokes flow simulations and compare these against their Poiseuille's law approximations based on the inlet and outlet width in the top row of Fig.~\ref{fig:diameter_effects_smooth}. The Poiseuille's law underestimates the flow conductance up to more than 70\%, due to the increase in the channel width inside the bifurcation. There are two reasons for the error: 1) the increase of bifurcation area; 2) the flow is no longer unidirectional in the bifurcation. 

To investigate the impact of the bifurcation area, we first connect the centre of three boundary curves with the bifurcation centre to separate the bifurcation into three regions with areas $A_0$, $A_1$ and $A_2$, as shown in Fig.~\ref{fig:smooth_example}(b). For each region, we keep the length at $L=2$ and scale the width of the corresponding channel to match the area ($\hat{D}=A_0/L$, $\hat{D}_1=A_1/L$ and $\hat{D}_2=A_2/L$), so that an area preserved conductance $\hat{G}$ can be calculated via Poiseuille's law. The relative difference between 2D Stokes flow simulations and area preserved conductances are presented in the bottom row of Fig.~\ref{fig:diameter_effects_smooth}. These results indicate that the failure of Poiseuille's law for predicting conductances in bifurcations with curved boundaries is primarily due to the disruption of the unidirectional flow assumption, instead of the change of bifurcation area.

\subsection{Effects of a fixed object on bifurcation flow}
\label{sec:object}

\begin{figure}
  \centering
    	\begin{subfigure}{0.49\textwidth}
		\centering
		\includegraphics[width=\textwidth]{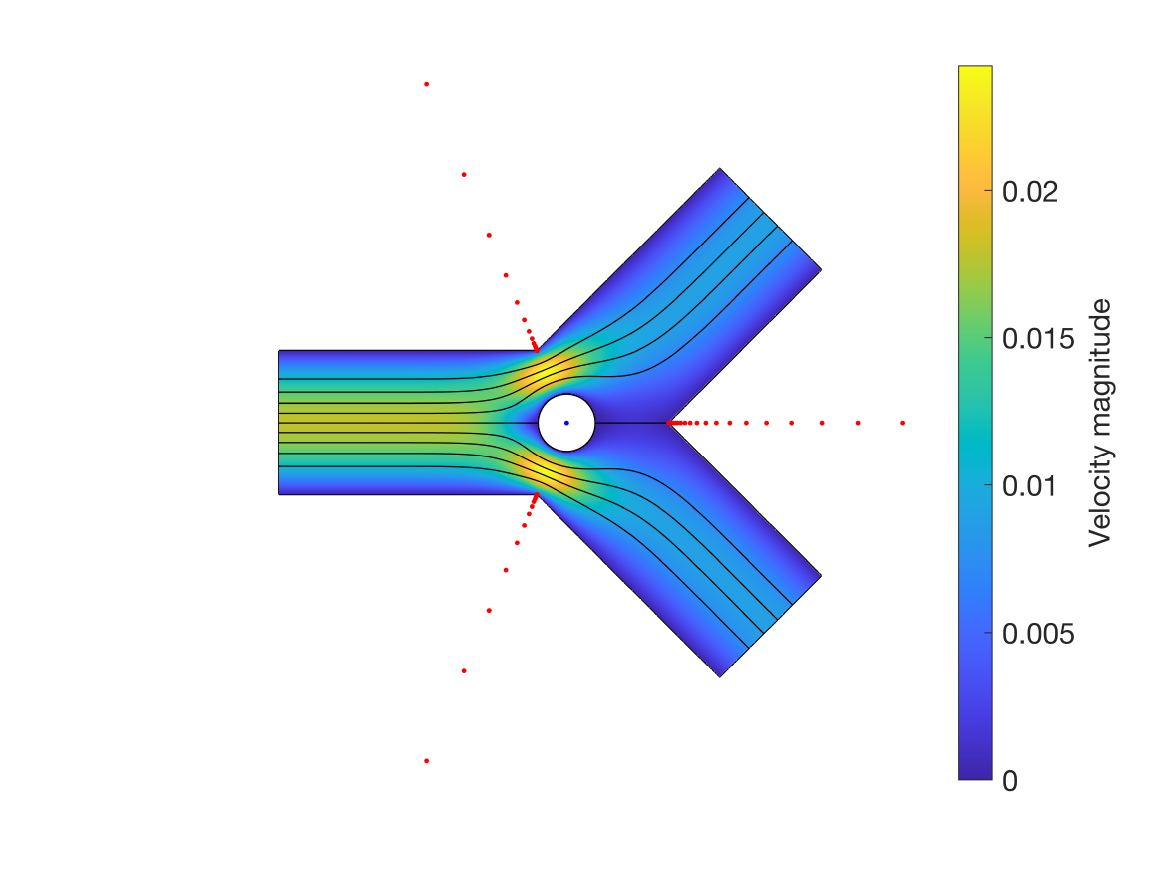}
		\caption{Simulation results}
	\end{subfigure}
	\begin{subfigure}{0.49\textwidth}
	    \centering
		\includegraphics[width=.5\textwidth]
  {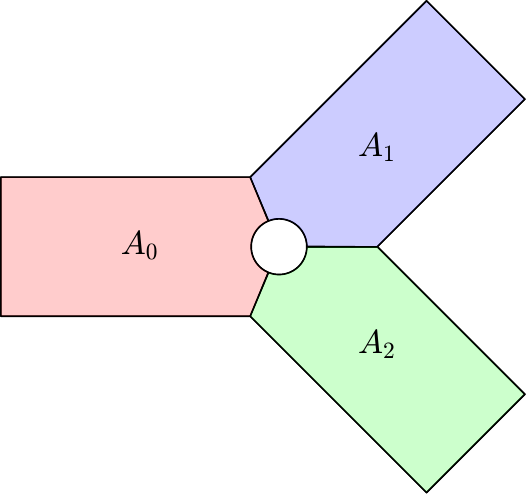}
  \vspace{30pt}
		\caption{Area for each channel}
	\end{subfigure}
  \caption{(a) Stokes flows in a 2D bifurcation with a fixed cylindrical particle at $(X_0,Y_0)=(0,0)$ and $R=0.2$ solved by the LARS algorithm, where $D_1=D_2=1$, $\alpha=\beta=\pi/4$, $L=2$ and $P_1=P_2=-1$. The locations of the poles and Laurent series are marked by red and blue dots, respectively. (b) Area for each channel when calculating area preserved conductance $\hat{G}$.}
  \label{fig:particle}
\end{figure}

We now consider bifurcations containing one fixed cylindrical particle in a symmetrical bifurcation ($D_1=D_2=1$, $\alpha=\beta=\pi/4$ and $L=2$) to investigate the effects of the location and size of the particle on the three flow conductances. Figure \ref{fig:particle}(a) shows the Stokes flows computed in a bifurcation for $D_1=D_2=1$, $\alpha=\beta=\pi/4$, $L=2$ and $P_1=P_2=-1$, with a fixed cylindrical particle at $(X_0,Y_0)=(0,0)$ and $R=0.2$. In addition to poles and a degree 80 polynomial, we also include a degree 20 Laurent series about the centre of the cylinder (blue point) in the rational function basis, as shown in Fig.~\ref{fig:particle}(a). The solution is computed to the accuracy $\mathcal{O}(10^{-7})$ in 0.6 second. A similar case has been considered in \cite{Xue2024}, where further details of the numerical algorithm can also be found.

\begin{figure}
  \centering
   \begin{subfigure}{\textwidth}
     \centering
     \includegraphics[width=.8\textwidth]{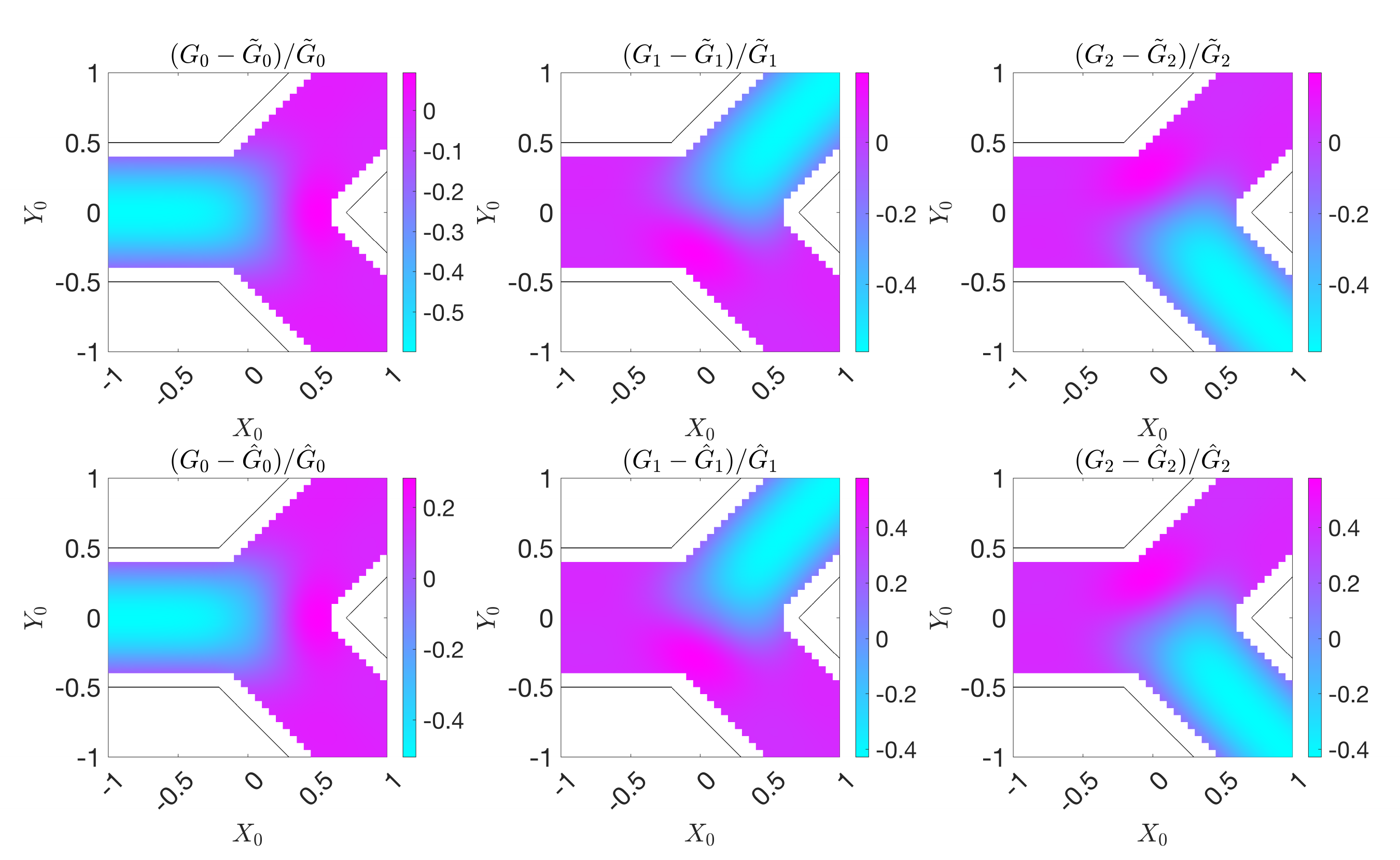}
     \caption{$R=0.1$}
 \end{subfigure}
  \begin{subfigure}{\textwidth}
     \centering
     \includegraphics[width=.8\textwidth]{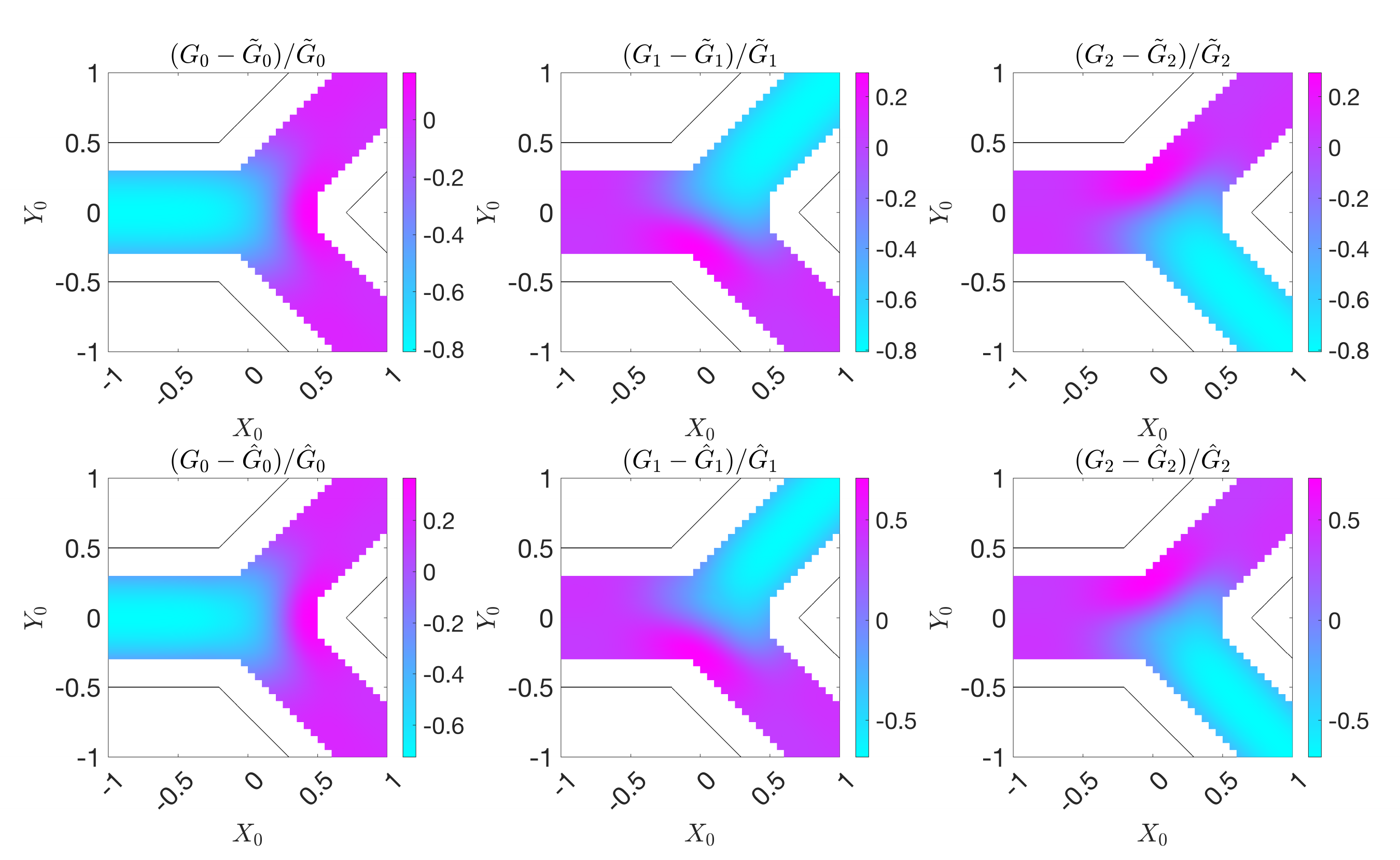}
     \caption{$R=0.2$}
 \end{subfigure}
 \caption{Flow conductances compared with Poiseuille's law approximation ($\tilde{G}$) and area preserved conductances ($\hat{G}$). The bifurcation has $D_1=D_2=1$, $\alpha=\beta=\pi/4$ and $L=2$ with a fixed cylindrical particle with different centre location $(X_0,Y_0)$ and radius $R$. The bifurcation boundaries are indicated by solid lines. The parameter space of $(X_0,Y_0)$ is chosen to prevent overlap between the particle and the channel, which results in a gap near the domain boundary.}
  \label{fig:particle_location}
\end{figure}

Figure \ref{fig:particle_location} shows the effects of $(X_0,Y_0)$ and $R$ on the flow conductance of three channels, compared with the Poiseuille's law approximation without considering the particle ($\tilde{G}$), and area preserved conductances by scaling the channel width to consider the area loss due to the presence of the particle ($\hat{G}$). Following the previous section, we connect the bifurcation centre and three bifurcating corners to separate the bifurcation into three regions, as shown in Fig.~\ref{fig:particle}(b), and compute the area for each region not occupied by the particle. For each region, we keep the length at $L=2$, and scale the width for the area preserved conductance calculation.

Similarly to the results from the previous section, the failure of Poiseuille's law for predicting conductances in bifurcations with a fixed particle is mainly due to the disruption of the unidirectional flow assumption, rather than the modifications to the bifurcation area due to the particle. We see that for the area preserved cases (second row of Fig.~\ref{fig:particle_location}(a)), the errors in $\hat{G}$ can be greater than 40\%. Note that the error in $\hat{G}$ can be greater than that in $\tilde{G}$ for channels without a particle, if one compares the upper limit of the colour scale between the first row and the second row of Fig.~\ref{fig:particle_location}(a), because the area preserved scaling leads to $\hat{D}<D$ and thus further underestimates the flow conductance. 

Furthermore, we consider the impact of particle sizes on flow conductance, where $R=0.1$ in Fig.~\ref{fig:particle_location}(a) and $R=0.2$ in Fig.~\ref{fig:particle_location}(b). When the particle radius doubles from 0.1 to 0.2, the flow conductance of the channel containing the particle is further reduced, resulting in increased errors in both $\tilde{G}$ and $\hat{G}$.

The presence of a fixed cylindrical particle mainly reduces $\tilde{G}$ of the channel containing the cylinder, while slightly affects other channels, as represented by the cyan regions in Fig.~\ref{fig:particle_location}. In addition, $\tilde{G}$ is higher when the gap between the cylinder and the channel wall is smaller. Following Williams et al.~\cite{Williams2020}, our results also suggest that the resistance of 2D channel (or bifurcation) flow is minimised when the particle touches the bifurcation boundary and thus has less disruption on the fast flows in the channel centre. This highlights the importance of considering the lateral location of the finite-sized objects using a 2D model for accurate flow conductance computations. 

\subsection{Learn the flow conductances using neural network models}
\label{sec:neural_network}
We see that the flow conductance is sensitive to bifurcation geometry and the presence of fixed finite-sized objects. We now determine the non-linear relationship between dimensionless geometric parameters and the conductance tensor using machine learning models. We perform 1000 simulations using the LARS algorithm in straight channel bifurcations for random geometrical parameters in $D_1,D_2\in[0.5,1]$, $\alpha,\beta\in[0,\pi/2]$ and $L=2$. Using the simulation results, we train three neural network models, which read inputs of $D_1$, $D_2$, $\alpha$ and $\beta$ and each generate a component of the conductance tensor $\mathbf{G}$. The neural network has a simple structure consisting of 3 fully-connected feedforward layers of size 20, and each layer uses a rectified linear unit (ReLU) activation function. We train the neural networks by minimising the mean squared error between the predicted and simulated conductance components using 80\% of the data, and validate the prediction using the other 20\% of the data. The training and testing were carried out using the {\tt fitrnet} function in MATLAB.

\begin{figure}
  \centering
   \begin{subfigure}{\textwidth}
     \centering
     \includegraphics[width=.75\textwidth]{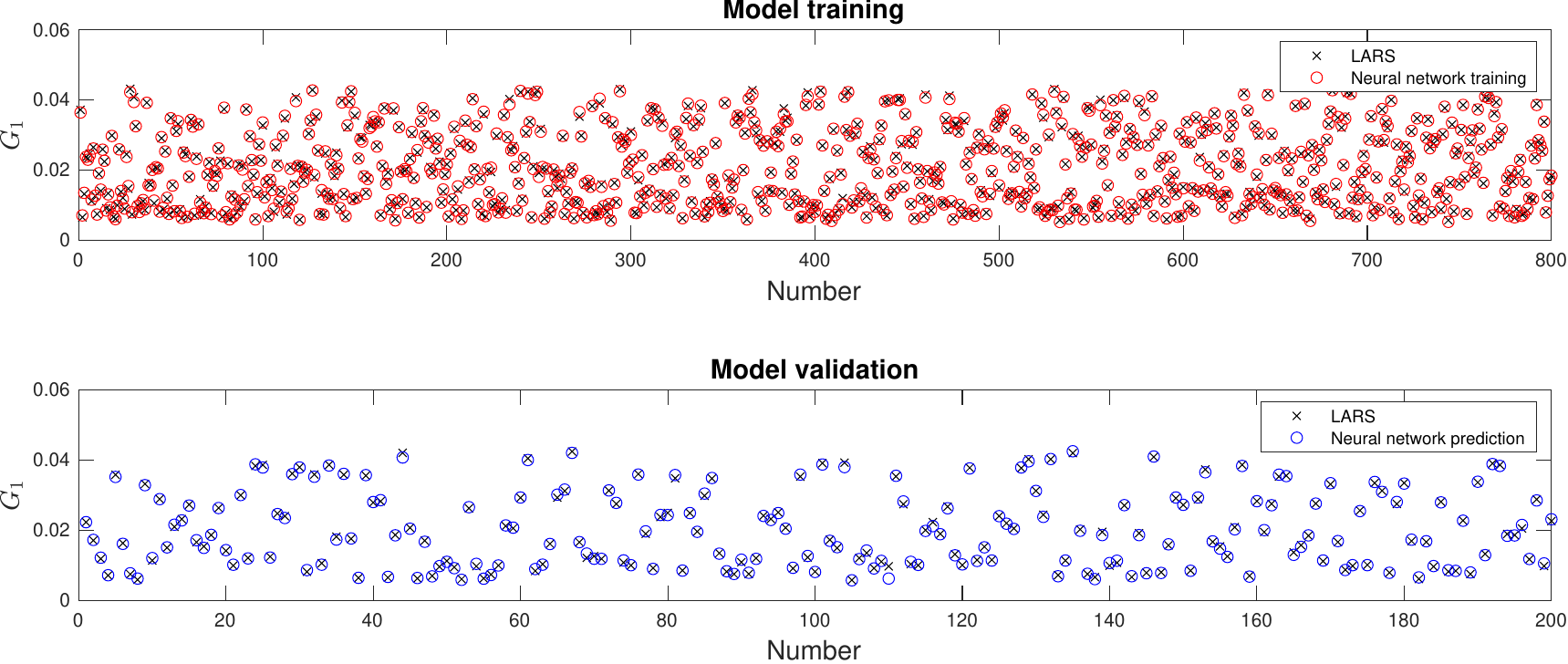}
     \caption{Bifurcations with straight channels}
 \end{subfigure}
  \begin{subfigure}{\textwidth}
     \centering
     \includegraphics[width=.75\textwidth]{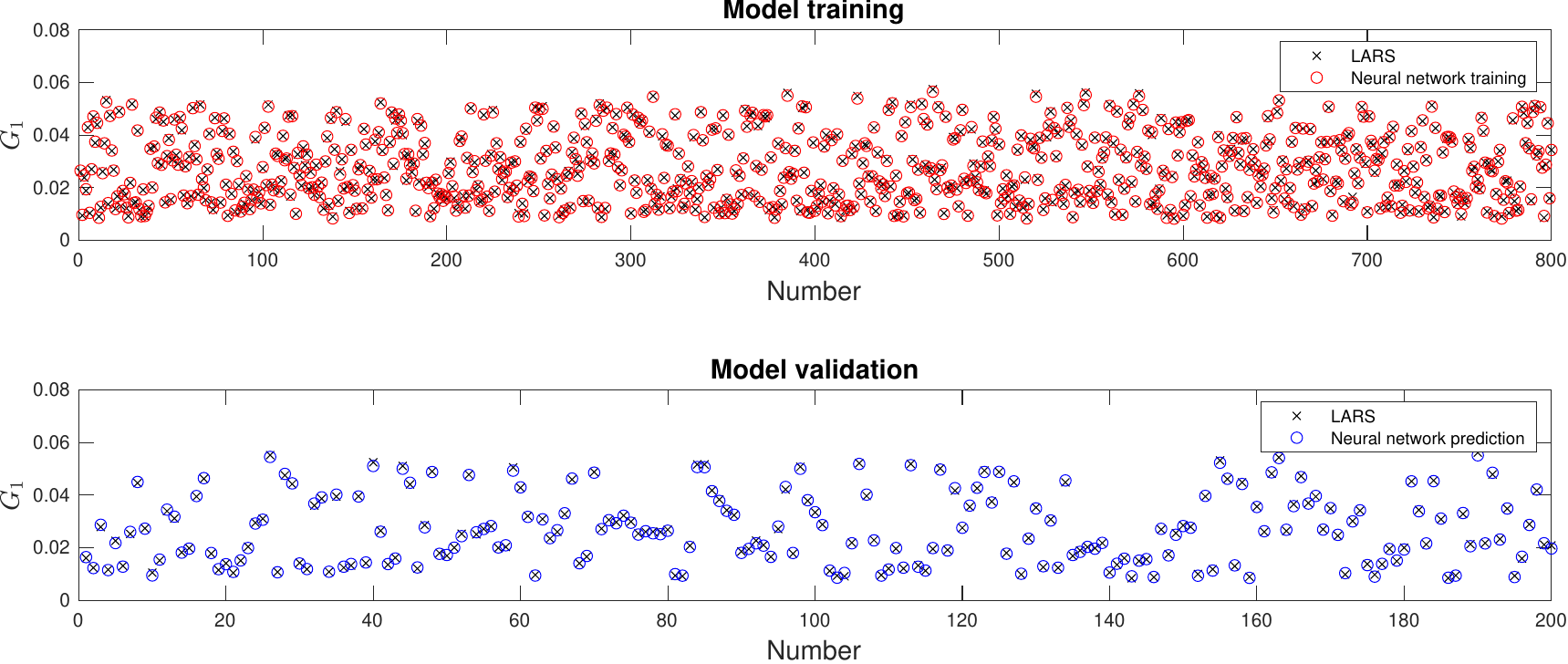}
     \caption{Bifurcations with curved boundaries}
 \end{subfigure}
   \begin{subfigure}{\textwidth}
     \centering
     \includegraphics[width=.75\textwidth]{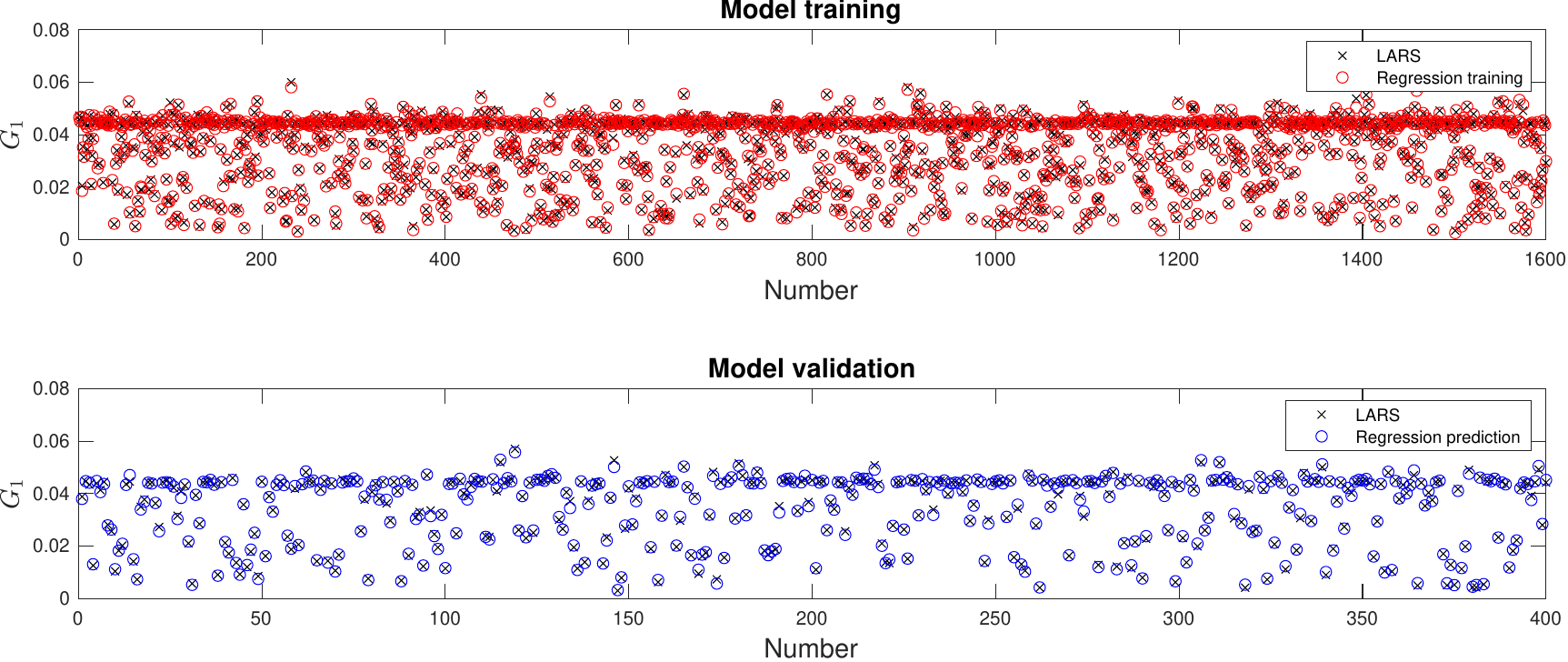}
     \caption{Bifurcation with a fixed cylindrical particle}
 \end{subfigure}
 \caption{Neural network training and validation for the conductance tensor for (a) bifurcations with straight channels; (b) bifurcations with curved boundaries; (c) bifurcation with a fixed cylindrical particle of different sizes at different locations. The LARS simulations and the neural network predictions are represented by crosses and circles, respectively. Only the results of $G_1$ are plotted, while all simulation results and models can be found in the GitHub repository provided in Appendix \ref{sec:appendix}.}
  \label{fig:neural_network}
\end{figure}

Figure \ref{fig:neural_network}a presents the comparison between the predicted and simulated $G_1$ of 1000 cases with different geometrical parameters. The top panel shows the training results of 800 cases, where the LARS simulation results and the neural network predictions are represented by crosses and red circles, respectively. The bottom panel validates the predictions for the remaining 200 cases using the trained neural network against the LARS simulations. The mean squared error between predicted and simulated $G_1$ in 200 validation cases is $2.27\times10^{-7}$. These results indicate that the flow conductance tensor for Stokes flows in any 2D bifurcation can be highly accurately approximated using its geometrical parameters and a neural network model learning their relationship from data.

In addition, we train neural networks in the same way for bifurcations with curved boundaries in the same parameter space of $D_1$, $D_2$, $\alpha$ and $\beta$, where the results of 1000 cases are presented in Fig.~\ref{fig:neural_network}b. Although the bifurcation geometry is now bounded by cubic B\'{e}zier curves instead of straight channels, the workflow provides a good approximation of conductance tensor. The mean squared error of $G_1$ in 200 validation cases is $2.29\times10^{-7}$.

In the previous section, we have shown that the presence of a fixed cylindrical particle can have a significant impact on flow conductances. In the same bifurcation as shown in Fig.~\ref{fig:particle}, we train neural networks to predict the conductance tensor as a function of the particle location ($X_0,Y_0\in[-1,1]$) and its radius ($R\in(0,0.3]$). For this scenario, we use 2000 LARS simulation results (1600 for training and 400 for validation). The mean squared error of $G_1$ in 400 validation cases is $5.15\times10^{-7}$. These results suggest the neural network approach is able to predict the flow conductance tensor for a bifurcation containing a particle of different size at different locations.

\subsection{Separation of Stokes flows in a 2D bifurcation for different outlet pressures}
\label{sec:streamline}

Having focused on computing flow conductances, we now investigate the streamline patterns in a bifurcation for different $P_1$ and $P_2$, which can only be understood via 2D Stokes flow models. Here we focus on the streamline that separates the fluids into two child channels. This is important, for example, when considering the advective transport of passive traces.

\begin{figure}
  \centering
  \includegraphics[width=.8\textwidth]{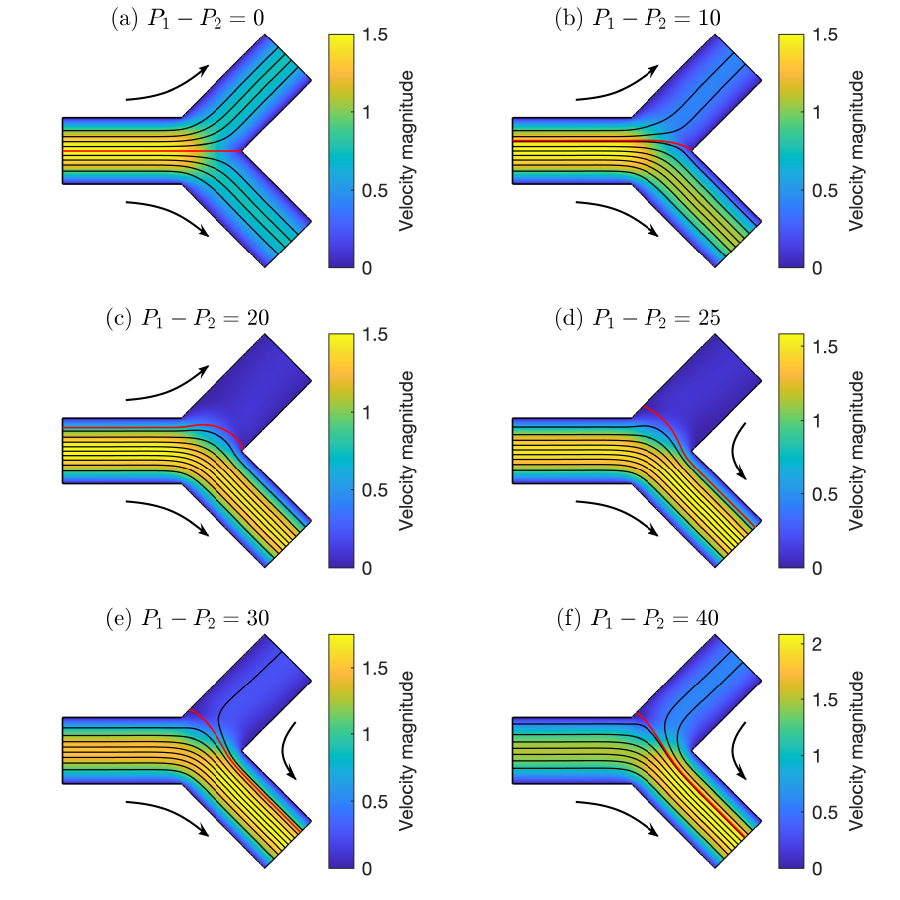}
  \caption{Stokes flows in a 2D bifurcation for different $P_1=P_2$, when $D_1=D_2=1$, $\alpha=\beta=\pi/4$, $L=2$ and $Q=1$. The streamline that separates the flows into two channels is coloured in red. Other streamlines are coloured in black. The flow directions are indicated using arrows outside.}
  \label{fig:pressure_cases}
\end{figure}

Figure \ref{fig:pressure_cases} displays the streamlines that separate the flows for different $P_1-P_2$, when $D_1=D_2=1$, $\alpha=\beta=\pi/4$ and $L=2$. In all scenarios, the pressures $P_1$ and $P_2$ are set to maintain a constant inlet flux $Q=1$. For cases (a)-(c), the red streamline separates the flow that enters two child channels. Note that the end point of the streamline is not exactly at (but very close to) the sharp corner between the two child channels in (b) and (c). For cases (d)-(f), the red streamline separates the flow from the parent channel and the first channel that enters the second channel, since the flow direction reverses in the first channel. Based on Stokes flow simulations, the reverse flow appears when $P_1-P_2>22.49$, while it is estimated to be $P_1-P_2=24$ using Poiseuille's law.

\begin{figure}
  \centering
  \includegraphics[width=.8\textwidth]{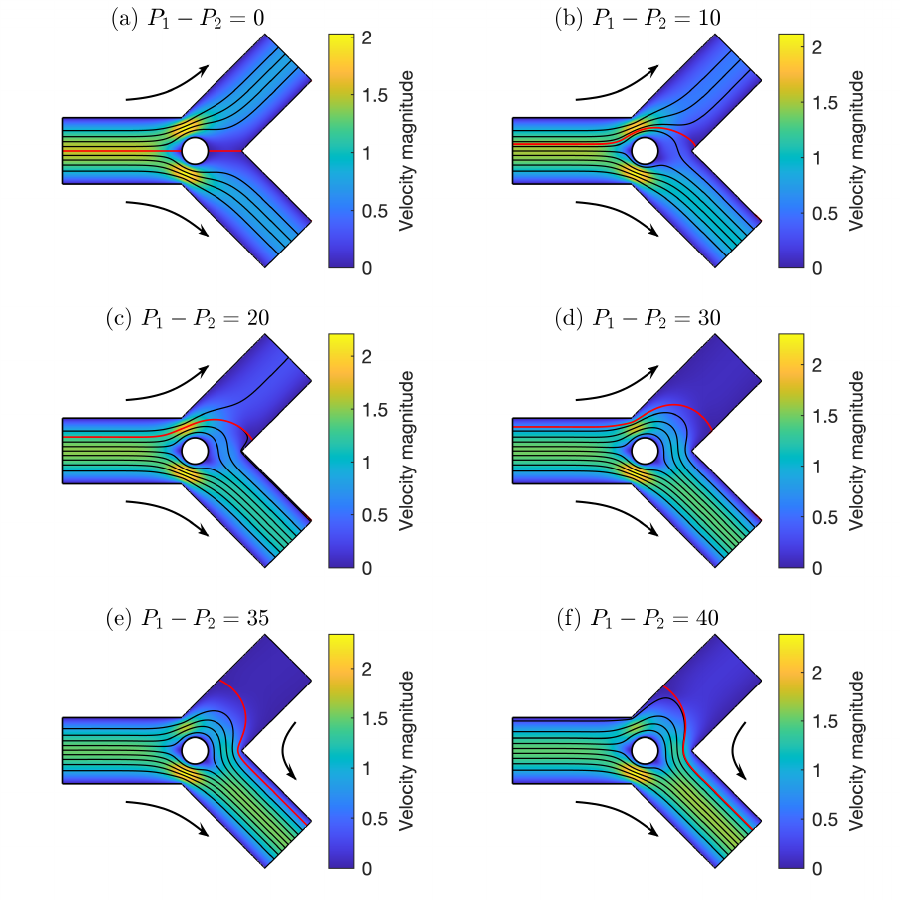}
  \caption{Stokes flows in a 2D bifurcation containing a fixed cylindrical particle ($(X_0,Y_0)=(0,0)$ and $R=0.2$) for different $P_1=P_2$, when $D_1=D_2=1$, $\alpha=\beta=\pi/4$, $L=2$ and $Q=1$. The streamline that separates the flows into two channels is coloured in red.}
  \label{fig:pressure_cases_particle}
\end{figure}

Figure \ref{fig:pressure_cases_particle} presents the separation of Stokes flows in the same geometry, but with a fixed cylindrical particle at the origin with a radius of 0.2. Compared with Fig.~\ref{fig:pressure_cases}, the reverse flow happens at approximately $P_1-P_2=33.22$. In addition, the end point of the streamline that separates the flows is much further away from the sharp corner between the two child channels, compared with the no-particle scenario.

\section{Discussion}
\label{sec:discussion}

In this paper, we computed 2D Stokes flows in a single bifurcation, and investigated the effects of bifurcation geometry on flows, with a focus on the flow conductances. Making full advantage of the great speed and accuracy of our 2D Stokes flow solver \cite{Xue2024}, we performed simulations for parameters including child channel widths and bifurcation angles. We considered the effects of domain boundaries and a fixed cylindrical cylinder on Stokes flows in a bifurcation. The relationship between the flow conductances and the geometrical parameters for these scenarios were learned via neural network models, and the accuracy of their predictions were validated against 2D Stokes simulation results. In addition, we investigated the separation of Stokes flows in a bifurcation for different outlet pressures. 

Our simulation results show that the bifurcation geometry can significantly impact the conductances. In Section \ref{sec:angle}, we present Stokes flows in bifurcations with the same channel widths, but different branching angles. All these bifurcations are represented by the same flow network model, if the flow conductances are approximated by Poiseuille's law. However, some of these approximations can have an error up to 9\% for $L=2$ in the parameter space of $\alpha$ and $\beta$ in Fig.~\ref{fig:angle_effects}. 

For bifurcation with curved boundaries instead of straight edges in Section \ref{sec:boundary_geometry}, the Poiseuille's law approximation significantly underestimates the conductances, mainly due to the disruption of the unidirectional flow assumption, while the increase of the bifurcation area plays a secondary role. These simulation results indicate Poiseuille's law is not suitable to approximate flow conductances of bifurcations with changing channel widths.

We also considered a fixed cylindrical particle in a bifurcation and its impact on conductance. Following the curved boundary scenarios, we show that the Poiseuille's law overestimates the channel conductance mainly due to the violation of the unidirectional flow assumption, rather than the loss of bifurcation area occupied by the particle. These results further show that Poiseuille's law is not able to predict accurate conductances in bifurcations with fixed finite-sized objects.

In addition, the particle mainly reduces the flow conductance of the channel containing the particle, while having little effect on other channels. This local effect holds even when the particle centre is within one channel width from the bifurcation centre (Fig.~\ref{fig:particle_location}). These results suggest, when computing the pressure-flux relationship in a large 2D flow network with multiple objects, we may identify channels containing objects, and only update the flow conductance of these channels, based on the local nature of the impact. Similar local effects are expected for 3D flows and moving objects, but further investigation will be required.

To improve the applicability of our simulation results to studies in flow networks, we fitted the flow conductances as machine learning models of dimensionless geometrical parameters describing the bifurcation. We show simple neural networks can be effective in representing accurate relationships between conductances and geometrical parameters. We provide our workflow and the trained neural network in an online repository, so now one can approximate the conductances in a 2D bifurcation without the need to run new simulations. Furthermore, it is possible to use the workflow to perform additional 2D Stokes flow simulations for additional bifurcation geometries using the LARS algorithm \cite{Xue2024}, and train new machine learning models.

These successful examples indicate that neural network models can provide a much more accurate approximation for flow conductances of bifurcations in a flow network than using Poiseuille's law. In a flow network that consists of several orders of bifurcations, one may decompose the network into units of segments and bifurcations, and use Poiseuille's law and trained neural networks to approximate their conductance, respectively.

We have shown that the neural network model is able to consider the impact of fixed cylindrical particles on conductances. To consider realistic scenarios of particle transport in a network, we need to investigate whether the machine learning models can accurately capture the impact of moving particles or multiple particles in a bifurcation. These problems involve more complex physics, necessitating the design of a neural network structure and optimisation procedure to accelerate the training and improve the accuracy of predictions \cite{LeCun2015,Raissi2019,Sirignano2023}. One recent example of applying machine learning techniques to predict the distribution of red blood cells in 3D microvascular networks can be found in \cite{Ebrahimi2022}.

It should be noted that the updated network model, despite considering bifurcation geometry details in approximating conductances, is still a 0D approach that compresses most flow information. The flow details, as presented in Section \ref{sec:streamline}, can only be obtained by Stokes flow simulations \cite{Xue2024}. In addition, it has been revealed in Section \ref{sec:object} that the flow conductance tensor of a bifurcation (or even a channel) with a fixed cylinder depends on the lateral location of the cylinder, another result that cannot be obtained from a 0D model.

The results presented in this work are purely 2D. In principle, a similar flow network for the pressure-flux relationship can be derived for any 3D bifurcation from 3D Stokes flow simulations. However, the parameter space that defines a 3D bifurcation is expected to be larger, requiring a more comprehensive search of parameter space, that might, for example, quantify out of plane effects when the three branches are not in the same plane.

In summary, we have demonstrated that incorporating detailed bifurcation geometry and fixed objects into flow network models significantly improves their accuracy in estimating flow conductances, using a workflow that combines the LARS algorithm and a machine learning approach. In addition, our simulation results have shown the limitation of Poiseuille's law approximations and underpinned the potential of machine learning models to provide more accurate predictions for flow and particle transport in branching networks.

\appendix
\section{MATLAB codes}
\label{sec:appendix}
All MATLAB codes and models can be found in a GitHub repository at \url{https://github.com/YidanXue/stokes_2d_bifurcation}.

\begin{acknowledgments}

YX would like to thank financial support from the UK EPSRC (EP/W522582/1). SJP is supported by a Yushan Fellowship from the Ministry of Education, Taiwan (111V1004-2). SLW and YX are grateful to funding from the UK MRC (MR/T015489/1).

\end{acknowledgments}

\bibliography{bibliography}

\end{document}